\documentclass[reprint,superscriptaddress,
amsmath,amssymb,pre]{revtex4-2}

\usepackage[utf8]{inputenc}
\usepackage{graphicx}
\usepackage{dcolumn}
\usepackage{bm}
\usepackage[hidelinks]{hyperref}
\usepackage{mathptmx}

\begin{document}

\title{Tensor network calculation of the logarithmic correction exponent in the XY model}

\author{Seongpyo Hong}

\affiliation{Department of Physics and Photon Science, Gwangju Institute of Science and Technology, Gwangju~61005, Korea}

\author{Dong-Hee Kim}
\email{dongheekim.gist.ac.kr}
\affiliation{Department of Physics and Photon Science, Gwangju Institute of Science and Technology, Gwangju~61005, Korea}
\affiliation{School of Physics, Korea Institute for Advanced Study, Seoul~02455, Korea}

\begin{abstract}
We study the logarithmic correction to the scaling of the first Lee-Yang (LY) zero in the classical $XY$ model on square lattices by using tensor renormalization group methods. In comparing the higher-order tensor renormalization group (HOTRG) and the loop-optimized tensor network renormalization (LoopTNR), we find that the entanglement filtering in LoopTNR is crucial to gaining high accuracy for the characterization of the logarithmic correction, while HOTRG still proposes approximate bounds for the zero location associated with two different bond-merging algorithms of the higher-order singular value decomposition and the oblique projectors. Using the LoopTNR data computed up to the system size of $L=1024$ in the $L \times L$ lattices, we estimate the logarithmic correction exponent $r = -0.0643(9)$ from the extrapolation of the finite-size effective exponent, which is comparable to the renormalization group prediction of $r = -1/16$.
\end{abstract}

\maketitle

\section{Introduction}
\label{sec:intro}

Multiplicative logarithmic corrections appear in the critical 
behaviors of certain statistical physics models, introducing 
another set of scaling exponents characterizing criticality
\cite{Kenna2006a,Kenna2006b,Kenna2012}.
In the Berezinskii-Kosterlitz-Thouless (BKT) transition
\cite{B1971,KT1972,KT1973,Kosterlitz1974},
the renormalization group (RG) equations predicted that
the correlation function $G(R)$ at the critical point
exhibits the leading-order behavior of 
$G(R) \sim R^{-\eta}(\ln R)^{-2r}$~\cite{Kosterlitz1974,Amit1980} 
or more generally
$G(R) \sim R^{-\eta}(b + \ln R)^{-2r}$~\cite{Kadanoff1981} 
with exponents $\eta = 1/4$ and $r = -1/16$
at a large distance $R$.
The logarithmic correction factor essentially
distinguishes the critical behaviors of the correlation 
function and susceptibility 
from those of the Ising model undergoing the second-order 
transition. 
On the other hand, it is numerically challenging to precisely 
identify such multiplicative logarithmic correction 
with a very small exponent. 
Much numerical effort has been devoted to 
measuring $r$ in the two-dimensional (2D) $XY$ model 
and related models undergoing the BKT transition
\cite{Kenna1995,Irving1996,Kenna1997,Patrascioiu1996,Campostrini1996,Janke1997,Jaster1998,Tomita2002,Chandrasekharan2003,Strouthos2004,Hasenbusch2005,Arisue2009,Komura2012}.
While early estimates of $r$ vary from positive 
to negative values (see Table~4 in Ref.~\cite{Kenna2006c}), 
later large-scale Monte Carlo (MC) simulations showed 
improved agreement with the RG prediction.

In previous MC studies of the 2D $XY$ model, 
Kenna and Irving~\cite{Kenna1995} firstly measured 
$r = -0.02(1)$ from the finite-size-scaling (FSS) analysis 
of the lowest lying (first) Lee-Yang (LY) zero 
for system sizes up to $L=256$ in the $L\times L$ 
square lattices.
At the critical point, they found that the first LY zero 
$\theta_1$ should behave with increasing system size $L$ as 
\begin{equation}
    \theta_1 \sim L^{\lambda} (\ln L)^r,
\end{equation}
which was derived from its relation to the leading-order 
scaling behavior of the susceptibility,
\begin{equation}
    \chi \sim L^{-d} \theta_1^{-2} \sim L^{2-\eta} (\ln L)^{-2r},
\end{equation}
where $d = 2$ for two dimensions and thus $\lambda = -2 + \eta/2$.
Using the Villain formulation, Janke~\cite{Janke1997} 
measured $r = -0.0270(10)$ from the FSS analysis 
of the susceptibility in the critical region for system sizes 
up to $L=512$.
Later, Hasenbusch~\cite{Hasenbusch2005} examined 
an alternative scaling ansatz of the susceptibility, 
\begin{equation}
    \chi \sim L^{2 - \eta}(C + \ln L)^{-2r} ,
\end{equation}
reporting $r = -0.056(7)$ from the FSS analysis with 
the MC dataset of $256 \le L \le 2048$ in the pure $XY$ model.
The high-temperature expansion 
done by Arisue \cite{Arisue2009} reported 
the similar value of $r = -0.054(10)$
from the calculation of the moments of the correlation function.
Most recently, Komura and Okabe \cite{Komura2012} performed 
large-scale MC calculations for sizes up to $L = 65536$,
reporting the best fit with $r = -0.064(4)$ 
at the fixed value of $C = \ln 16$ in the FSS analysis of 
the susceptibility. The parameter $C$ effectively includes 
subleading-order corrections that may decay rather slowly
with increasing $L$ \cite{Hasenbusch2005}.
Setting $C=0$ provided smaller values of $r = -0.0406(3)$ 
in Ref.~\cite{Hasenbusch2005} and 
$r \approx -0.55$ in Ref.~\cite{Komura2012}
with similar system sizes. 

In this paper, we revisit the FSS analysis of the first
LY zero in the 2D $XY$ model by employing 
methods based on the tensor renormalization group (TRG).
Since the first MC measurement of the LY zero~\cite{Kenna1995},
there have been no other attempts to measure 
the logarithmic correction exponent 
using the LY zero in the $XY$ model.
Most of other previous estimates of $r$ were based on 
the susceptibility that might have been more 
straightforwardly measurable in cluster MC simulations.
The purpose of the present work is to examine applicability
of the TRG-based methods to the numerical identification
of the first LY zero and then to provide an updated estimate of 
the logarithmic correction exponent.

The TRG methods provide a deterministic way
of evaluating the partition function of a classical spin model
in the tensor network representation
\cite{TRG}. 
The higher-order tensor renormalization 
group (HOTRG) method~\cite{HOTRG} was previously 
applied to the Fisher zero problem where 
the partition function is evaluated 
at a complex temperature \cite{Denbleyker2014,Hong2020}. 
In the Ising and Potts models, the HOTRG method
was also used to obtain the density of the LY zeros 
from the discontinuity of magnetization \cite{GarciaSaez2015}. 
While tensor network methods have been actively
applied to study phase transitions in classical and 
quantum systems~\cite{Okunishi2022}, including 
the BKT transitions~\cite{Yu2014,Chatelain2014,Chen2017,Chen2018,Li2020,Huang2020,Ueda2020,Ueda2021},
the computation of the first LY zero 
in the $XY$ model has not been studied with TRG yet. 
It still remains unclear whether or not 
a TRG-based method such as HOTRG allows enough accuracy 
to characterize such delicate logarithmic correction
with a small exponent predicted at the BKT transition. 

We compare HOTRG with the loop-optimized tensor 
network renormalization (LoopTNR) \cite{LoopTNR} 
in identifying the location of the first LY zero at the critical point.
It turns out that HOTRG fails to give a converged estimate 
at a large system,
although it still proposes approximate bounds 
for the zero location that are set by
the estimates associated with two 
different bond-merging algorithms based on
the higher-order singular value decomposition \cite{HOTRG} 
and the oblique projector method \cite{Iino2019}.
In contrast, the LoopTNR calculations 
show much better convergence with increasing 
the bond dimension cutoff, indicating the importance of 
removing the short-range entanglement~\cite{TEFR}. 
We obtain the first LY zeros for system sizes up to 
$L=1024$ in the $XY$ model. 
In the analysis of the alternative scaling ansatz with
an undetermined constant as being introduced 
in the susceptibility~\cite{Hasenbusch2005}, 
we present that our finite-size estimate of 
the logarithmic correction exponent 
approaches closer to the RG prediction, 
providing the updated estimate
of $r = -0.0643(9)$ from extrapolation.

This paper is organized as follows. 
In Sec.~\ref{sec:procedures},
we describe the numerical procedures including
a brief review of HOTRG and the two bond-merging algorithms
and the performance of our initial state preparation 
for the loop optimization in the LoopTNR calculations. 
In Sec.~\ref{sec:result},
we present the comparison between the two bond-merging
algorithms of HOTRG and the estimate with LoopTNR 
in computing the first LY zero and the analysis of 
the LoopTNR data of the LY zero to measure
the logarithmic correction exponent in the $XY$ model.
Summary and conclusions are given in Sec.~\ref{sec:summary}.

\section{Numerical Procedures}
\label{sec:procedures}

\subsection{XY model and Lee-Yang zeros}

The classical $XY$ model is described by the Hamiltonian,
\begin{equation} \label{eq:H}
    H = - J \sum_{\langle i,j \rangle} \cos (\phi_i-\phi_j)
        -h \sum_i \cos \phi_i ,
\end{equation}
where $\phi_i$ is a spin angle at site $i$, and $h$ denotes 
a magnetic field. The coupling strength $J$ and the Boltzmann
factor $k_\mathrm{B}$ are set to be unity, and thus
the temperature unit $J/k_\mathrm{B}$ and 
the magnetic field unit $J$ are omitted for brevity 
throughout this paper. We consider the periodic boundary 
conditions.

The zeros of a partition function provide an alternative 
tool to study phase transitions and critical phenomena
(see, for instance, Refs.~\cite{Bena2005,Janke2001} and references therein). 
The LY zeros~\cite{YL1952,LY1952} are defined in the plane of complex fugacity
while the Fisher zeros~\cite{FisherZero} are defined 
in the plane of complex temperature.
Characterizing the BKT transition using partition function
zeros has been interest of many previous works 
\cite{Kenna1995,Irving1996,Kenna1997,Janke2002,Hwang2009,Denbleyker2014,Rocha2016,Costa2017,Kim2017,Hong2020}.
In the models that satisfy the Lee-Yang theorem~\cite{LY1952}, 
including the $XY$ model~\cite{Dunlop1975},
the LY zeros are exactly on the imaginary axis of
the magnetic field.
The first LY zero is the one with the smallest magnitude, 
exhibiting a characteristic scaling behavior 
with increasing system size at the critical point.

Finding the location of the LY zero requires the evaluation 
of the partition function $Z = \sum_{\{\phi_i\}} \exp(-\beta H)$
at an imaginary magnetic field $h = i \theta$.
In the $XY$ model, we fix the inverse temperature $\beta$
at the critical point $\beta_c = 1.1199$
that is agreed between the previous large-scale Monte Carlo
\cite{Hasenbusch2005,Komura2012},
high-temperature expansion \cite{Arisue2009}, 
and tensor network renormalization \cite{Ueda2021} studies.
To identify the first LY zero $\theta_1$, we first graphically 
locate an approximate location of $\theta_1$ and 
then numerically minimize $|Z(\beta_c, \theta)|$ 
to refine the estimate of $\theta_1$.

In the TRG formulation~\cite{TRG},
the partition function of a classical spin model with 
local interactions is written in the square lattices as
\begin{equation} \label{eq:Zexact}
    Z(\beta_c, \theta) = \mathrm{Tr} 
    \prod_i T_{x_i x'_i y_i y'_i},
\end{equation}
where $T$ is a local tensor, and its four legs are
associated with the bonds in the $x$ and $y$ directions.
In the $XY$ model, the local tensor is given as
\cite{Liu2013,Yu2014}
\begin{equation} \label{eq:localT}
    T_{x x' y y'} = 
    \sqrt{I_x(\beta_c) I_{x'}(\beta_c) 
    I_y(\beta_c) I_{y'}(\beta_c)} \,
    I_{x+y-x'-y'}(i\beta_c \theta) ,
\end{equation}
where $I_n$ is the modified Bessel function of the first kind.
While the exact enumeration of the tensor product 
in Eq.~(\ref{eq:Zexact}) is numerically impossible unless
the system is very small, 
TRG provides a controlled way to compute 
$Z$ by coarse-graining the tensor network
with bond dimension truncation.
Below we briefly describe the procedures of 
the two TRG-based methods that we employ 
to evaluate the partition function.

\subsection{HOTRG and bond-merging methods}
    
\begin{figure}[t]
    \centering
    \includegraphics[width=0.47\textwidth]{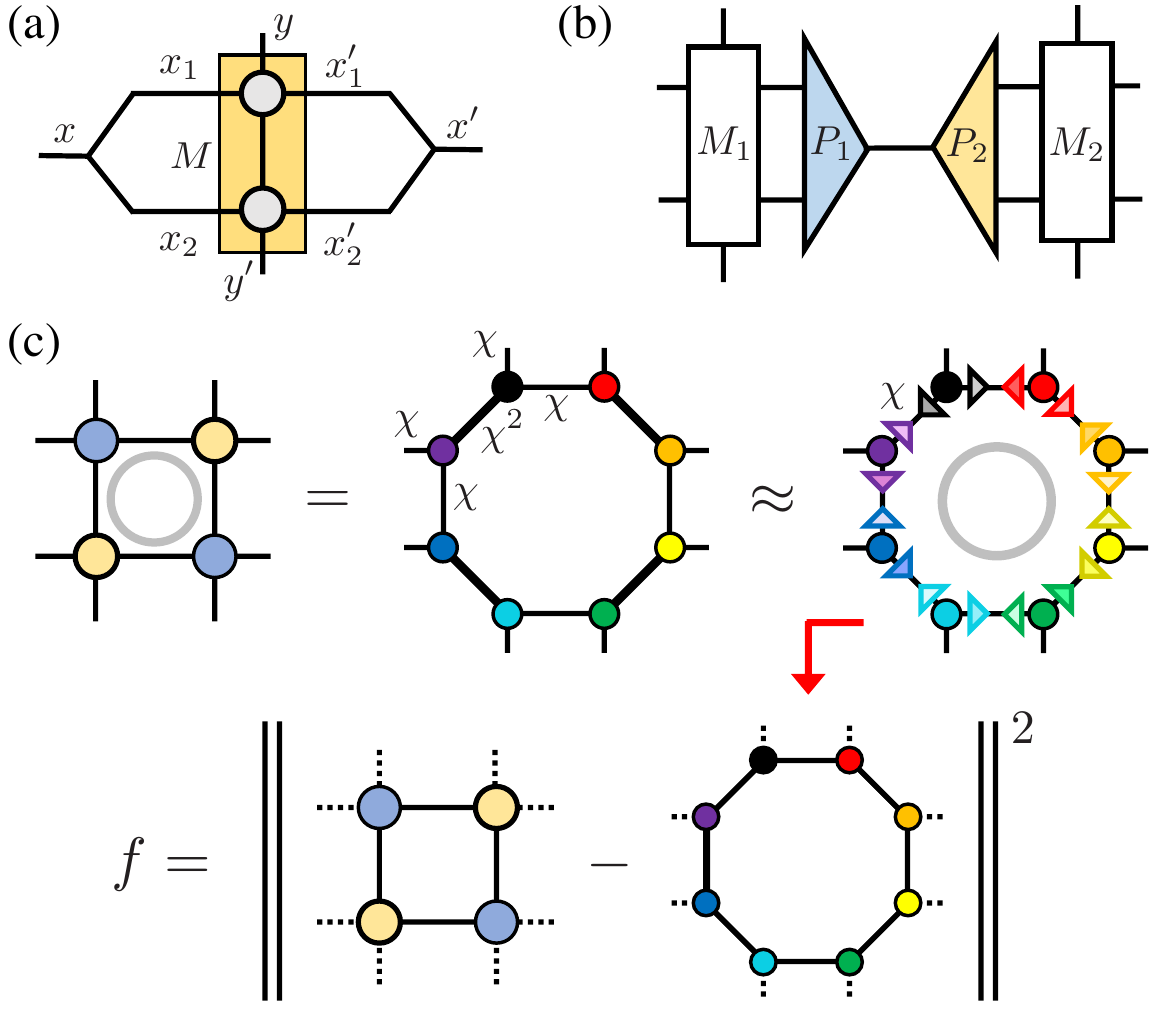}
    \caption{Schematic diagram of (a) the combine tensor $M$ 
    in HOTRG, (b) the oblique projectors, and (c) the initial
    tensor preparation for the loop optimization in LoopTNR.
    In (c), thicker bonds have dimension as large 
    as $\chi^2$ that is to be truncated by the projectors 
    of the entanglement filtering.
    }
    \label{fig1}
\end{figure}

An essential part of the HOTRG procedures is the step of 
merging a pair of parallel bonds into a one bond
in the contraction of two neighboring tensors.
In the square lattices of $2^N \times 2^N$ sites, 
with translation invariance being imposed, 
the final coarse-grained tensor is obtained by performing
$2N$ contractions alternatively along the $x$ 
and $y$ directions.
For instance, as shown in Fig.~\ref{fig1}(a), 
one can write down the contraction of 
two neighboring tensors along the $y$ direction as
\begin{equation}
    M_{x x' y y'} = \sum_i T_{x_1 x'_1 y i} T_{x_2 x'_2 i y'},
\end{equation}
where the bond dimension of $x \equiv x_1 \otimes x_2$ and 
$x' \equiv x'_1 \otimes x'_2$ increases to $\chi^2$ 
if each leg of $T$ has dimension $\chi$. 
The crucial part of HOTRG is to keep the dimension 
of $x$ and $x'$ below a numerically manageable 
cutoff $\chi$. The truncation error is due to 
the finite cutoff limited by available 
computing resources.

The original HOTRG paper \cite{HOTRG} proposed 
the higher-order singular value decomposition (HOSVD) 
for the truncation as
\begin{equation}
    T'_{x x' y y'} = \sum_{ij} U_{ix} M_{i j y y'} U_{jx'}^*, 
\end{equation}
where the matrix $U$ is determined by 
solving an eigenproblem of $MM^\dagger$. 
To preserve the symmetry of the local tensor
at an imaginary magnetic field, we perform the orthogonal
transformation by diagonalizing the real part of $MM^\dagger$ 
in the same way that was used in 
Refs.~\cite{Denbleyker2014,Hong2020} for the Fisher zero
problem.

The other bond-merging algorithm~\cite{Iino2019} considers 
a pair of the oblique projectors $P_1$ and $P_2$
inserted between the neighboring combined tensors $M_1$ and $M_2$,
as sketched in Fig.~\ref{fig1}(b),
minimizing $|| M_1 M_2 - M_1 P_1 P_2 M_2 ||$ 
at a given cutoff $\chi$ of the bond dimension between them.
While details of the algorithm can be found  
in the literature~\cite{Iino2019,Yoshiyama2020,Morita2021}, 
let us briefly review the numerical procedures.
The projectors are given as 
\begin{eqnarray}
P_1 &=& R_2 \tilde{V}_t \tilde{\Sigma}_t^{-1/2}, \\
P_2 &=& \tilde{\Sigma}_t^{-1/2} \tilde{U}_t^\dagger R_1 ,
\end{eqnarray}
where $R_1$ and $R_2$ are from the QR and RQ factorization
of $M_1$ and $M_2$, respectively, and the other tensors 
are from the truncated singular value decomposition (SVD) of 
$R_1 R_2 \approx \tilde{U}_t \tilde{\Sigma}_t \tilde{V}_t^\dagger$
that keeps the largest $\chi$ singular values.
The $R$ tensors can be computed using matrix diagonalization as
\begin{eqnarray}
R_1 &=& \Lambda_1^{1/2} U_1 , \\
R_2 &=& U_2^\dagger \Lambda_2^{1/2} , 
\end{eqnarray}
where $M_1^\dagger M_1 = U_1^\dagger \Lambda_1 U_1$ 
and $M_2 M_2^\dagger = U_2^\dagger \Lambda_2 U_2$.
Finally, the contraction along the $y$ direction 
is done as
\begin{equation}
    T'_{x x' y y'} = \sum_{ij} [P_2]_{xi} M_{ijyy'} [P_1]_{jx'}.
\end{equation}
While the symmetry of the local tensor in Eq.~(\ref{eq:localT}) 
is not explicitly preserved with the oblique projectors 
at a complex field, the first LY zero computed using 
the projectors shows accuracy comparable 
to the symmetry-preserved HOSVD. These two bond-merging
algorithms play complementary roles
in the search for the LY zero location. It turns out that
they provides approximate upper and lower bounds for 
the true zero location, which we will demonstrate later in Sec.~\ref{sec:result}.

\subsection{Loop optimization of tensor network renormalization}

A known issue of TRG is that it does not make an isolated 
RG flow because of the survival of the short-range entanglement
\cite{TEFR}. HOTRG is much more accurate than the original 
TRG at a non-critical region, but it also suffers from 
the same issue of the original TRG at a critical point, 
which may cause inaccuracy in finding the LY zero 
especially at a large system.
Several methods~\cite{TEFR,EV-TNR,Evenbly2017,LoopTNR,TNRplus,Gilt-TNR} 
have been proposed to remove the short-range 
entanglement and demonstrated that 
a correct fixed point tensor is recovered 
with much higher accuracy at a critical point. 
So far, the effect of the entanglement filtering 
remains untested in the plane of a complex field 
or temperature for a partition function zero problem.
In Sec.~\ref{sec:result}, we will show that removing
the short-range entanglement is crucial 
particularly to the identification of the multiplication 
logarithmic correction in the $XY$ model.

\begin{figure}[t]
    \centering
    \includegraphics[width=0.47\textwidth]{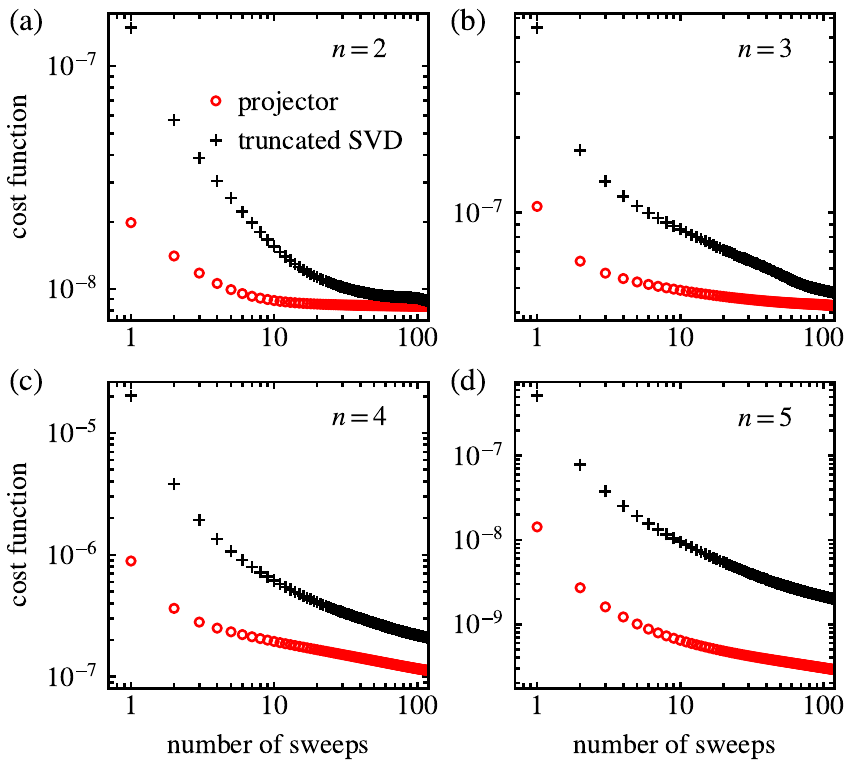}
    \caption{Convergence of the loop optimization.
    The cost function $f$ is plotted as a function 
    of the number of sweeps in the loop optimization tested 
    at the $n$-th coarse-graining step in the 2D $XY$ model. 
    The marker with ``projector'' represents our initial tensor 
    preparation method where the truncation occurs 
    with the entanglement filtering. The other marker  
    represents the initial tensor prepared by the truncated SVD
    of the TRG scheme. The calculations are done at 
    $\beta = \beta_c$ and $h = i$ with the bond dimension 
    cutoff $\chi = 40$.}
    \label{fig2}
\end{figure}

We adopt the LoopTNR method~\cite{LoopTNR} that extends 
the TRG scheme by adding the entanglement filtering step 
to remove the corner double line tensors and 
replacing the truncated SVD of the original TRG with 
the loop optimization. 
We have implemented our code by faithfully following the original
paper~\cite{LoopTNR} yet with extra care of preparing an initial 
tensor for the loop optimization.
It was already pointed out in the original paper 
that choosing a good initial tensor could considerably speed up 
the convergence of iterations in the loop optimization.

The simplest way of preparing an initial octagonal tensor ring 
for the loop optimization is to perform the truncated SVD
as done in the original TRG scheme. 
Instead, as sketched in Fig.~\ref{fig1}(c), 
we use the entanglement filtering algorithm to 
generate projectors to truncate the bond dimension.
We first apply SVD to the local tensors with all singular values
being kept and then perform the entanglement filtering 
on the eight-tensor ring. In the final step, the projectors
are constructed by choosing the largest $\chi$ singular values.
During the loop optimization, the entanglement 
filtering is performed every ten sweeps for better stability.
The maximum number of sweep is limited to $200$.
Figure~\ref{fig2} presents comparison between
two choices of the initial tensors, showing that 
the entanglement filtering projectors gains 
order-of-magnitude improvement over the simple truncated SVD
in the minimization of the cost function.

\section{Results and Discussion}
\label{sec:result}

\subsection{Comparison between HOTRG and LoopTNR}

\begin{figure}
    \centering
    \includegraphics[width=0.47\textwidth]{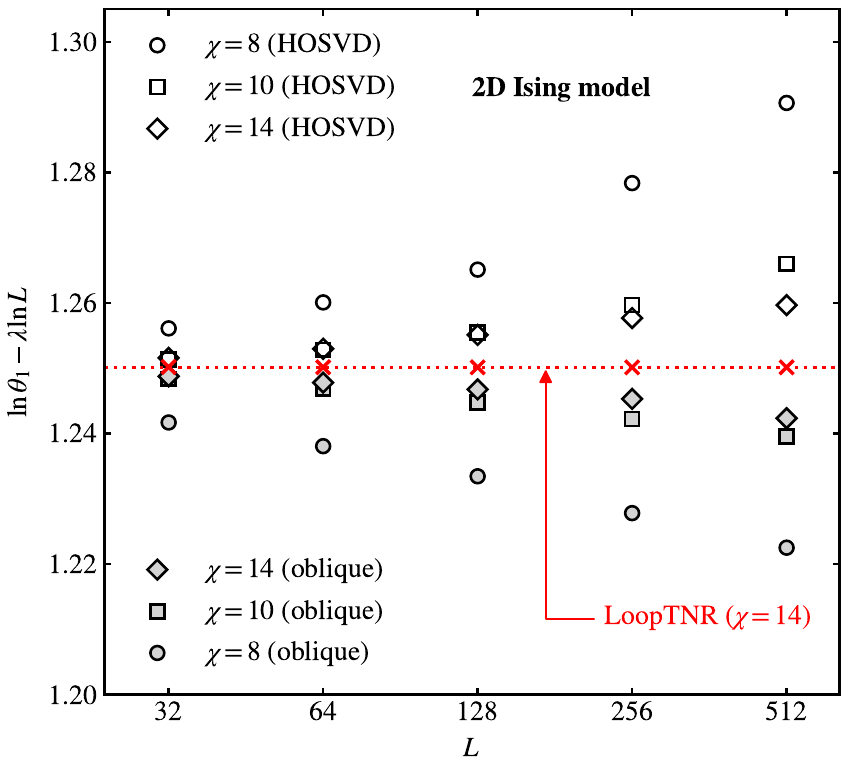}
    \caption{Comparison between HOTRG and 
    LoopTNR in finding the first LY zero of 
    the 2D Ising model at the critical point. 
    Different cutoff values of the bond dimension 
    ($\chi = 8, 10, 14$) are examined with
    the HOSVD and oblique projector methods in 
    the bond-merging step of HOTRG. 
    The exponent $\lambda$ is fixed at $-15/8$. 
    The dotted line indicates the exact scaling behavior.}
    \label{fig3}
\end{figure}

We compare the cutoff dimension dependence of 
the LY zero estimates computed using HOTRG and LoopTNR.
Our findings in the following are based on 
common observations in the Ising and $XY$ models.
First, in the HOTRG calculations, two bond-merging 
algorithms approach each other from the opposite sides 
as the cutoff $\chi$ increases. While it fails to converge,
the comparison between the estimates 
associated with the two bond-merging algorithms proposes 
the upper and lower bounds for the zero location.
Second, LoopTNR converges much faster than HOTRG
and thus provides a more reliable estimate of the zero
location. The comparison between HOTRG and LoopTNR 
shows the importance of the entanglement filtering 
to the precise identification of the LY zeros in the $XY$ model.

\begin{figure}
    \centering
    \includegraphics[width=0.47\textwidth]{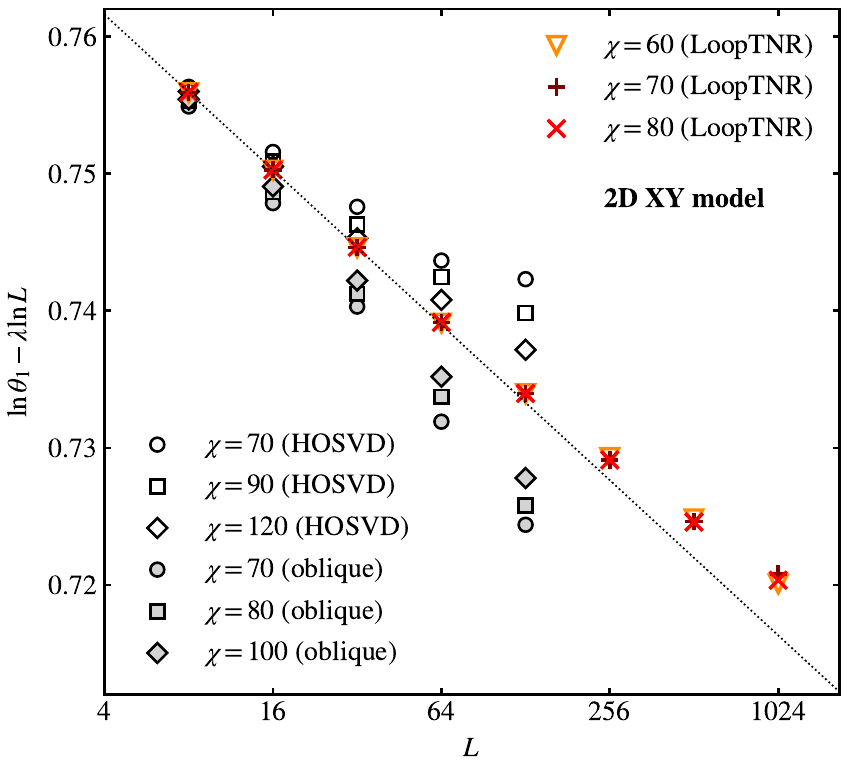}
    \caption{First LY zero of the 2D $XY$ model 
    at the critical point $\beta_c = 1.1199$. 
    The LoopTNR results are compared with the HOTRG 
    estimates based on the HOSVD and oblique projector methods.
    The exponent $\lambda$ is fixed at $-15/8$.
    The dotted line of $L^{-0.0082}$ is given for comparison 
    with a pure power law.
    }
    \label{fig4}
\end{figure}

Figure~\ref{fig3} displays the first LY zeros computed 
in the 2D Ising model. The LoopTNR calculations verify 
the exact scaling behavior $\theta_1(L) \propto L^\lambda$
with the critical exponent $\lambda = -15/8$ 
at a relatively low cutoff $\chi = 14$. 
On the other hand, the HOTRG calculations converge rather 
slowly with increasing $\chi$, which gets worse 
as it goes to larger systems. Interestingly 
in the HOTRG calculations, the direction of the LY zeros 
moving toward the exact scaling line with increasing $\chi$ 
depends on which bond-merging algorithm is used. 
While the one with HOSVD approaches
the exact scaling line of the first LY zero from above,
the other with the oblique projectors lies below the exact line,
proposing an area where the exact LY zero should be located. 
Although our observation is purely 
empirical, testing the different bond-merging algorithms
may help judging the absolute convergence 
of the LY zero estimate.
In the Ising model, one can simply increase 
$\chi$ to see that the two HOTRG estimates 
indeed meet each other on the exact scaling line 
for the system sizes shown in Fig.~\ref{fig3}. 

Figure~\ref{fig4} shows the same tendency with 
the bond-merging algorithms of the HOTRG estimates 
in the $XY$ model.
The zero estimate moves with increasing $\chi$ 
from the opposite directions associated with 
the two bond-merging algorithms.
The situation in the $XY$ model is in fact much worse 
than in the Ising model. We fail to make these 
two HOTRG estimates meet together for $L \ge 16$ 
even at the largest cutoff that we have examined.
Thus, it is not practically possible to study 
the multiplicative logarithmic correction 
to the scaling of the LY zero 
by using the HOTRG calculations.

In contrast, the estimates of the LY zeros 
from the LoopTNR calculations with the cutoffs 
of $\chi = 60, 70, 80$ graphically overlap onto each other. 
All are well within the bounds proposed by the HOTRG estimates.
While the convergence with different $\chi$'s is
not perfect for the largest $L$ as seen 
in the numeric data of the LY zeros listed in Table~\ref{tab1},
the larger bond dimension makes the less truncation error in 
building a coarse-grained tensor and should provide 
the more accurate data. 
Our measurements of the exponent $r$ of 
the multiplicative logarithmic correction to the scaling 
presented below are
mainly based on the data of the largest cutoff $\chi = 80$
that we have managed to reach in our LoopTNR calculations.

\begin{figure}
    \centering
    \includegraphics[width=0.47\textwidth]{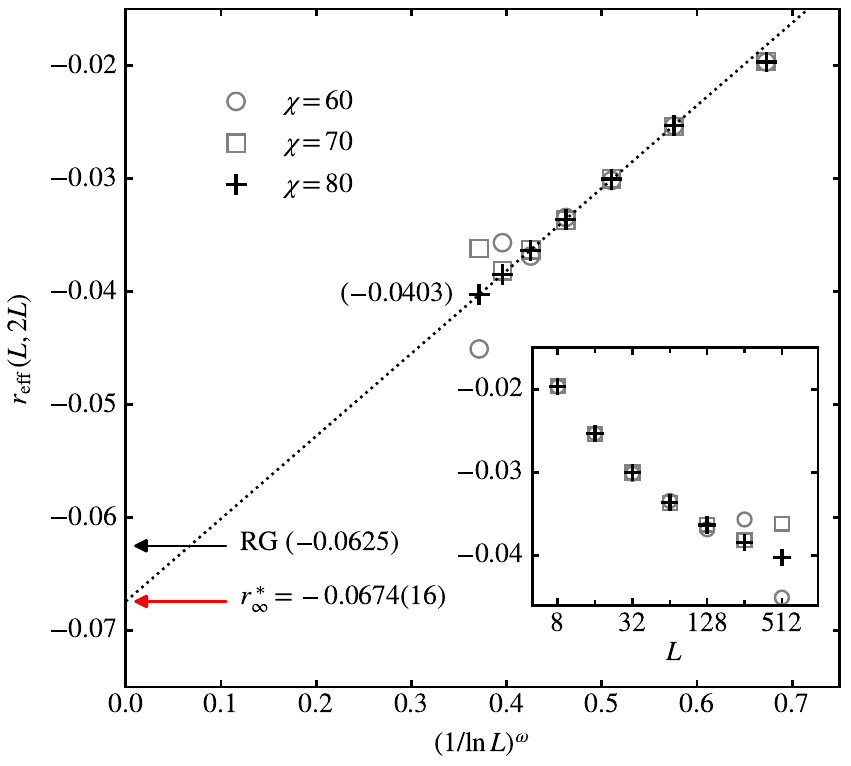}
    \caption{Logarithmic correction exponent estimate based on 
    the scaling ansatz of $\theta_1 \propto L^\lambda (\ln L)^r$.
    The exponent $\lambda$ is fixed at $-15/8$.
    The finite-size exponent $r_\mathrm{eff}$ 
    is computed for every consecutive sizes of $(L, 2L)$ 
    using the LoopTNR data of $\chi = 60, 70, 80$. 
    The extrapolation along the line of 
    $r_\mathrm{eff}(L) = r^*_\infty + a (\ln L)^{-\omega}$
    is shown at the parameter $\omega = 0.541$ 
    obtained from a fit to the data points of $\chi = 80$ 
    with $L = 8$ being excluded.
    }
    \label{fig5}
\end{figure}

\subsection{Logarithmic correction exponent}

We measure the logarithmic correction exponent
$r$ by examining two possible forms of the FSS ansatz.
First we examine the ansatz of the asymptotic scaling behavior, 
\begin{equation} \label{eq:ansatz1}
    \theta_1(L) \sim L^\lambda (\ln L)^r .
\end{equation}
which is the same one considered in the previous MC
study of the LY zero in the $XY$ model \cite{Kenna1995}. 
In finite-size systems, there must be an influence 
from non-universal subleading-order terms that decay with 
increasing $L$. This finite-size effect
is expected to be particularly problematic when trying 
to identify the logarithmic correction exponent 
because its base is the logarithm of the system size.
An ideal FSS analysis to determine the exponent $r$ 
would need a dataset of very large system sizes, 
such as a series of $\log_2 L = 2^n$, to perform 
a conventional log-log fit.
However, the sizes allowed in our calculations are 
$l \equiv \log_2 L = 3, 4, 5, \ldots, 10$,
implying that a significant finite-size effect could
appear in the evaluation of the exponent.

\begin{figure}
    \centering
    \includegraphics[width=0.47\textwidth]{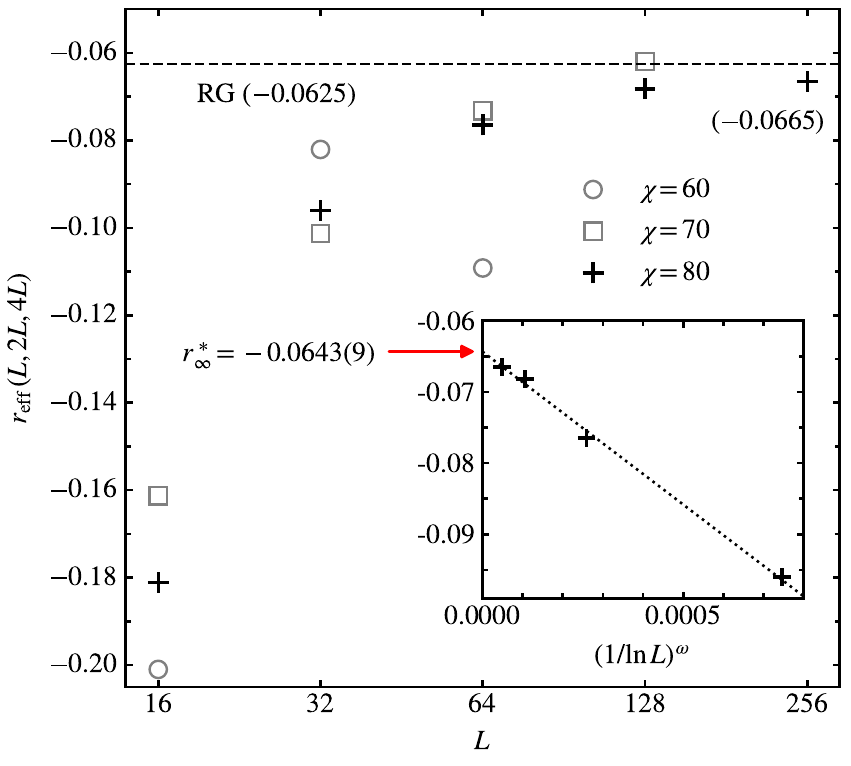}
    \caption{Logarithmic correction exponent estimate based on
    the scaling ansatz of $\theta_1 \propto L^\lambda (C + \ln L)^r$.
    The exponent $\lambda$ is fixed at $-15/8$.
    The finite-size logarithmic correction exponent $r_\mathrm{eff}$ 
    is computed with three system sizes of $(L, 2L, 4L)$ 
    in the LoopTNR data of $\chi = 80$. 
    The extrapolation along the line of 
    $r_\mathrm{eff}(L, 2L, 4L) = r^*_\infty + a (\ln L)^{-\omega}$
    is shown at the fitting parameter of $\omega = 5.79$ 
    obtained with the data point of $L = 8$ being excluded.
    }
    \label{fig6}
\end{figure}

The LoopTNR calculation is deterministic at a given 
bond dimension cutoff $\chi$ and free from a stochastic 
uncertainty. 
If the LoopTNR data of the first LY zero is precise enough,
a good way to get an exponent can be extrapolating
the finite-size exponents that are computed by equating $r$ 
in Eq.~(\ref{eq:ansatz1}) with two different system sizes.
which is extrapolated toward the wanted exponent $r$ 
in the thermodynamic limit. 
This method is sensitive to the accuracy of 
the raw data and thus typically used for numerically 
exact data at small systems. 
We write the finite-size exponent $r_\mathrm{eff}$ 
by choosing two consecutive system sizes of $l$ and $l+1$ as
\begin{equation} \label{eq:reff1}
    r_\mathrm{eff}(l) = 
    \left(\log_2\frac{\theta_1(l+1)}{\theta_1(l)} 
    - \lambda\right)
    \bigg/ \log_2\left(1+\frac{1}{l}\right),
\end{equation}
where $\lambda$ is fixed at $-15/8$. 
Supposed no numerical error existing in $\theta_1(l)$, 
the finite-size behavior of $r_\mathrm{eff}(l)$ is 
solely due to the subleading-order contributions that 
decrease with increasing $l$, implying that $r_\mathrm{eff}$
monotonically approaches the exact value of 
the exponent $r$ as $l$ increases.

Figure~\ref{fig5} presents $r_\mathrm{eff}(l)$ obtained
from the two-point estimate of Eq.~(\ref{eq:reff1}).
It turns out that the data with the largest cutoff available 
($\chi = 80$) shows a smooth monotonic curve 
expected in this method while the less accurate ones 
with the lower cutoffs indicate deviations from the one 
with the largest cutoff at $L = 512$ and $1024$.
We perform the extrapolation with the data of $\chi = 80$
along the model line of
\begin{equation} \label{eq:extrapol}
    r_\mathrm{eff}(l) = r^*_\infty + a_o l^{-\omega},
\end{equation}
finding $r^*_\infty = - 0.0674(16)$. Although this 
number is close to the RG prediction $r_\mathrm{RG} = -0.0625$, 
we must point out the risks of such extrapolation.
This extrapolation model assumes decay in the form of 
$l^{-\omega}$, which lacks a theoretical ground.
In addition, the available data points are quite far 
from the intercept at $1/l = 0$ on the extrapolation line, 
implying that the intercept $r$ may significantly 
vary with a choice of the model.
While this issue is fundamental, we argue that it is less 
severe if we take the alternative ansatz 
with an undetermined constant.

\begin{table}
    \begin{ruledtabular}
    \begin{tabular}{lccc}
    $L$ & $\chi=60$ & $\chi=70$ & $\chi=80$ \\ 
    \hline
    8    &   0.043152022481
         &   0.043152341820   
         &   0.043152516125  \\	
    16   &   0.011697992784   
         &   0.011698219091   
         &   0.011698141443  \\
    32   &   0.003171206193   
         &   0.003171237666   
         &   0.003171264869  \\ 
    64   &   0.000859814590 
         &   0.000859847214  
         &   0.000859849347  \\			
    128  &   0.000233203560   
         &   0.000233203767   
         &   0.000233206505  \\			
    256  &   0.000063265430  
         &   0.000063270187   
         &   0.000063270583  \\
    512  &   0.000017175547
         &   0.000017171787  
         &   0.000017171295  \\
    1024 &   0.000004660332  
         &   0.000004663680   
         &   0.000004661544  \\ 
    \end{tabular}
    \end{ruledtabular}
    \caption{Numeric data of the first LY zero 
    computed using the LoopTNR method with the bond dimension
    cutoff $\chi = 60, 70, 80$ in the 2D $XY$ model.}
    \label{tab1}
\end{table}

In the FSS analysis of the BKT transition, 
the logarithmic correction is often described by 
$(C + \ln L)$ with an undetermined constant $C$
instead of $\ln L$ \cite{Weber1988,Hasenbusch2005}.
Similarly, we may write the system-size scaling ansatz of 
the first LY zero as 
\begin{equation} \label{eq:ansatz2}
    \theta_1(L) \sim L^\lambda (C + \ln L)^r,
\end{equation}
where the constant $C$ may help us to include 
some of subleading-order contributions within the ansatz.
To determine the two unknowns of $r_\mathrm{eff}$ and $C$, 
we need to consider three system sizes $(l, l+1, l+2)$. 
The equation for $c \equiv C / \ln 2$ is then written as
\begin{equation} \label{eq:c}
    \frac{\log_2\frac{\theta_1(l+1)}{\theta_1(l)}-\lambda}
    {\log_2\frac{\theta_1(l+2)}{\theta_1(l+1)}-\lambda}
    = 
    \frac{\log_2\left[1+(c+l)^{-1}\right]}
    {\log_2\left[1+(c+l+1)^{-1}\right]} ,
\end{equation}
which is to be solved numerically. The right-hand side
is a bounded and monotonic function of $c$, indicating
that there exists a single solution or no solution.
Once $c$ is determined, the finite-size exponent 
$r_\mathrm{eff}$ can be computed as
\begin{equation} \label{eq:reff2}
    r_\mathrm{eff}(l) =
    \left(\log_2\frac{\theta_1(l+1)}{\theta_1(l)} 
    - \lambda\right)
    \bigg/ \log_2\left(1+\frac{1}{c+l}\right).
\end{equation}
Figure~\ref{fig6} shows $r_\mathrm{eff}$ obtained from 
the three-point analysis of Eqs.~(\ref{eq:c}) and (\ref{eq:reff2}). 
As one may have expected already in the two-point analysis, 
the datasets of $\chi = 60, 70$ having larger truncation 
errors do not give a solution of $c$ at large $L$'s. 
In contrast, the dataset of $\chi = 80$
provides a solution for $c$ at all $l$'s examined.
The three-point estimate of $r_\mathrm{eff}$ 
seems to saturate faster as $l$ increases than the two-point 
estimate without $C$, which may be seen from the value of 
$\omega = 5.7$ in the extrapolation along the line of
Eq.~(\ref{eq:extrapol}), while it was $\omega = 0.541$ 
in the case of the two-point analysis.
The intercept $r^*_\infty = -0.0643(9)$ is close to 
the data of $r_\mathrm{eff} = -0.0665$ obtained 
at the largest $l$ and is also well compared to 
the RG prediction. 

Finally, it is also worth to note that our dataset
proposes a possible range of $r$ without 
extrapolation being attempted.
As displayed in Figs. \ref{fig5} and \ref{fig6}, 
the two-point estimate of $r_\mathrm{eff}$ 
monotonically decreases with increasing $l$ while 
the three-point estimate increases with $l$.
The two curves have to meet at a true value of $r$ 
in the limit of infinite $l$.
Therefore, these monotonic yet contrasting behaviors of
$r_\mathrm{eff}(l)$ observed in the two different analysis
indicate that the exact $r$ must be in the range of 
$-0.0665 < r < -0.0403$,
where the numbers are given by $r_\mathrm{eff}$ 
at the largest $l$ available in Figs. \ref{fig5} and 
\ref{fig6}.
This range is probably the most conservative measure of
$r$ that we can provide, although the power-law extrapolation 
suggests that the true $r$ is likely to be much closer 
to the lower bound of the range.

\section{Summary and Conclusions}
\label{sec:summary}

We have investigated the applicability of the TRG-based
methods of HOTRG and LoopTNR to the calculation of 
the first LY zero, with a particular focus 
on the multiplicative logarithmic correction to 
the scaling at the critical point in the 2D $XY$ model.
It turns out that while LoopTNR exhibits graphical 
convergence in the LY zero location, HOTRG fails
to provide a reliable estimate within our accessible 
bond dimension cutoffs, assuring the importance
of the entanglement filtering in LoopTNR.
Despite the failure of HOTRG, we have found that 
the opposite convergence directions associated with
the two different bond-merging algorithms of HOSVD 
and the oblique projectors can propose the bounds for 
the zero location between which the LoopTNR estimate 
indeed resides. 

By using the LoopTNR dataset of the first LY zeros,
we have measured the logarithmic correction exponent 
$r$ in the $XY$ model. 
We have considered the finite-size effective exponent 
$r_\mathrm{eff}$ that is computed from a adjacent set of 
of the LY zero data.
In the two-point and three-point analysis of $r_\mathrm{eff}$ 
based on the two types of a scaling ansatz, we have identified 
the range of $-0.0665 < r < -0.0403$ 
at the largest system size examined for the measure of
the exponent.
The extrapolation with the three-point estimates provides
$r_\infty^* = -0.0643(9)$ that is well compared to 
the RG-predicted value of $r = -0.0625$.

Our estimates of $r$ are based on the LoopTNR dataset 
of $\chi = 80$ that is the largest cutoff accessible within
our computing resources. 
The irregularity observed at the lower values of the cutoff
implies that $\chi = 80$ may be the minimum 
bond dimension cutoff for LoopTNR to achieve enough 
accuracy required to the proper characterization of 
the multiplicative logarithmic correction.
While it is computationally challenging to increase $\chi$
further larger in the present study,
our three-point analysis method of computing $r_\mathrm{eff}$ 
based on the ansatz of Eq.~(\ref{eq:ansatz2}) can be 
a benchmark for future tensor network calculations 
at larger values of $\chi$ to pursue a more precise 
measurement of the logarithmic correction exponent 
in the $XY$ model and other systems undergoing 
the BKT transition.

\begin{acknowledgments}
The authors are grateful to Katharine Hyatt for guidance 
in using the GPU-accelerated ITensor~\cite{itensor} library.  
This work was supported from the Basic Science Research Program 
through the National Research Foundation of Korea
funded by the Ministry of Science and ICT (NRF-2019R1F1A106321).
Computing resources are provided by the KISTI supercomputing center
(KSC-2021-CRE-0165).
\end{acknowledgments}

\bibliographystyle{apsrev4-2} 
\bibliography{refs}

\end{document}